\documentclass[twocolumn,preprintnumbers,superscriptaddress]{revtex4}%
\usepackage{amssymb}
\usepackage{amsmath}
\usepackage{graphicx}
\usepackage{epstopdf}
\usepackage{dcolumn}
\usepackage{bm}
\usepackage{braket}
\usepackage{amsfonts}
\usepackage{xcolor}
\usepackage{soul}
\usepackage{hyperref}%
\hypersetup{
	colorlinks=true,
	linkcolor=blue,
	filecolor=magenta,      
	urlcolor=green,
}
\usepackage{ifthen}
\providecommand{\U}[1]{\protect\rule{.1in}{.1in}}
\providecommand{\U}[1]{\protect\rule{.1in}{.1in}}
\DeclareMathOperator{\Tr}{Tr}

\def\showal{1} 
\newcommand{\al}[1]{\ifthenelse{\showal=1}{\textcolor{orange}{[[#1]]}}{}}

\newcommand{\eb}[1]{\ifthenelse{\showal=1}{\textcolor{cyan}{[[#1]]}}{}}

\begin{document}
\title{Driving-induced resonance narrowing in a strongly coupled cavity-qubit system}
\author{Eyal Buks}
\thanks{These two authors contributed equally.}
\affiliation{Andrew and Erna Viterbi Department of Electrical Engineering, Technion, Haifa 32000 Israel}
\author{Paul Brookes}
\thanks{These two authors contributed equally.}
\affiliation{Department of Physics and Astronomy,  University College London,
Gower  Street,  London,  WC1E  6BT,  United  Kingdom }
\author{Eran Ginossar}
\affiliation{Advanced  Technology  Institute  and  Department  of  Physics,
University  of  Surrey,  Guildford,  GU2  7XH,  United  Kingdom }
\author{Chunqing Deng}
\affiliation{Institute for Quantum Computing, University of Waterloo, Waterloo, Ontario, Canada N2L 3G1}
\affiliation{Current address: Alibaba Quantum Laboratory, Alibaba Group, Hangzhou, Zhejiang 311121, P.R.China}
\author{Jean-Luc F. X. Orgiazzi}
\affiliation{Institute for Quantum Computing, University of Waterloo, Waterloo, Ontario, Canada N2L 3G1}
\author{Martin Otto}
\affiliation{Institute for Quantum Computing, University of Waterloo, Waterloo, Ontario, Canada N2L 3G1}
\author{ Adrian Lupascu }
\affiliation{Institute for Quantum Computing, University of Waterloo, Waterloo, Ontario, Canada N2L 3G1}

\date{\today }

	\begin{abstract}
		We study a system consisting of a superconducting flux qubit strongly coupled to a microwave cavity. Externally applied qubit driving is employed in order to manipulate the spectrum of dressed states. We observe resonance narrowing in the region where the splitting between the two dressed fundamental resonances is tuned to zero. The narrowing in this region of overlapping resonances can be exploited for long-time storage of quantum states. In addition, we measure the response to strong cavity mode driving, and find a qualitative deviation between the experimental results and the predictions of a semiclassical model. On the other hand, good agreement is obtained using theoretical predictions obtained by numerically integrating the master equation governing the system's dynamics. The observed response demonstrates a process of a coherent cancellation of two meta-stable dressed states.
		
	\end{abstract}
	\maketitle
	
	
	\section{Introduction}
	
	The spectral response of a variety of both classical and quantum systems near an isolated resonance is often well-described by the Breit-Wigner model \cite{Breit_519}. In this description the lifetime of an isolated resonance can be determined from its linewidth. A variety of intriguing effects may occur in regions where resonances overlap \cite{Fano_1866}. For example, both linewidth narrowing and broadening have been observed with systems having overlapping resonances \cite{Devdariani_477}. These effects are attributed to interference between different processes contributing to damping \cite{Dittes_215,Friedrich_3231}. Destructive interference gives rise to linewidth narrowing, whereas the opposite effect of broadening occurs due to constructive interference.
	
	\begin{figure}
		[ptb]
		\begin{center}
			\includegraphics[
			height=2.5949in,
			width=3.4537in
			]%
			{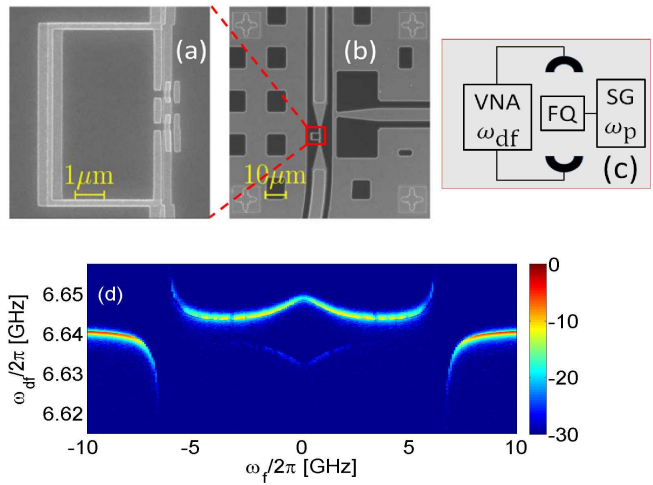}%
			\caption{The device. (a) Electron micrograph of the flux qubit. (b) Zoom out
				electron micrograph showing the qubit embedded in the CPW resonator and its
				local flux control line. (c) Sketch of the experimental setup. The cavity transmission is measured using a vector network analyzer (VNA). Monochromatic flux qubit (FQ) driving is applied using a signal generator (SG). (d) The measured cavity transmission (in dB units)
				vs. $\omega_{\mathrm{f}}/2\pi$ (magnetic field detuning from the symmetry
				point) and $\omega_{\mathrm{df}}/2\pi$ (cavity driving frequency). The power
				injected into the cavity is $-112$ dBm. For the device under study
				$\omega_{\mathrm{c}}/2\pi=6.6408\operatorname{GHz}$, $\omega_{\Delta}%
				/2\pi=1.12\operatorname{GHz}$, $g/2\pi=0.274\operatorname{GHz}$ and
				$\gamma_{\mathrm{c}}/\omega_{\mathrm{c}}=1.1\times10^{-5}$. The relaxation
				time $T_{1}=1.2\operatorname{\mu s}\left(  1+0.45\operatorname{ns}%
				\times\left\vert \omega_{\mathrm{f}}\right\vert \right)  $ is obtained from
				energy relaxation measurements, and the rate $T_{2}^{-1}=4.5\operatorname{MHz}%
				\left(  1+44\left\vert \omega_{\mathrm{f}}\right\vert /\omega_{\mathrm{a}%
				}\right)  $ is obtained from Ramsey rate measurements
				\cite{Orgiazzi_1407_1346}. The empirical expressions for both $T_{1}$ and $T_{2}$ are obtained using approximate interpolation.}%
			\label{FigNA}%
		\end{center}
	\end{figure}
	
	These effects have been demonstrated in a wide variety of both classical and quantum systems. In the classical domain narrowing has been observed with resonators having two overlapping resonances for which the frequency separation is smaller than the resonances' bandwidth \cite{Braginsky_9906108,Xu_123901,Liu_789}. Closely related processes occur in the quantum domain with systems having overlapping resonances. In some cases this overlap is obtained by static tuning of the system under study. One well-known example is the Purcell effect \cite{Purcell_839}, which is observed when atoms interact with light confined inside a cavity. In such cavity quantum electrodynamics (CQED) systems, both linewidth narrowing and broadening occur when the atomic and cavity mode resonances overlap. Other examples of static tuning giving rise to linewidth narrowing and broadening due to overlapping resonances have been reported in \cite{Makhmetov_247,Seipp_1}. 
	
	Closely related processes occur in atomic systems exhibiting electromagnetically induced transparency (EIT) \cite{Marangos_471,Wu_053806}. However, tuning into the region of EIT is commonly based on external driving (rather than static tuning), which can be used for manipulating the spectrum of the dressed states. Both linewidth narrowing and broadening have been observed in such systems in the region where the dressed spectrum contains overlapping resonances. Commonly, a broadened resonance is referred to as a bright state, whereas the term dark state refers to a narrowed resonance. The slow propagation speed associated with dark states \cite{Budker_1767} can be exploited for long term storage of quantum information \cite{Fleischhauer_022314}.
	
	Here we report on a linewidth narrowing that is experimentally observed in a superconducting circuit composed of a microwave resonator and a Josephson flux qubit \cite{Mooij_1036,Orlando_15398}. The qubit under study, which is strongly coupled \cite{Wallraff_162,Niemczyk_772,Forn_237001,Forn_1804_09275} to a coplanar waveguide (CPW) microwave resonator \cite{Niemczyk_772,Abdumalikov_180502,Bal_1324,Orgiazzi_104518,Jerger_042604,Oelsner_172505,Inomata_140508}, is shown in Fig.~\ref{FigNA}(a) and (b). The strong coupling gives rise to a dispersive splitting of the cavity mode resonance.  We find that this frequency splitting can be controlled by applying a monochromatic driving to the flux qubit [see Fig.~\ref{FigNA}(b)]. The effect of linewidth narrowing, which is discussed below in section \ref{Sec_QD}, is observed when the frequency and power of qubit driving are tuned into the region where the frequency splitting vanishes. In this region the measured linewidth becomes significantly smaller than the linewidth of the decoupled cavity resonance by a factor of up to 20.
	
	While the linewidth narrowing effect is induced by qubit driving, a variety of other nonlinear effects can be observed with strong cavity mode driving \cite{Serban_022305,Laflamme_033803,Siddiqi_207002,Lupacscu_127003,Boaknin_0702445,Mallet_791,Boissonneault_060305,Boissonneault_100504,Boissonneault_022324,Boissonneault_022305,Boissonneault_013819,Reed_173601,Ong_047001,Ong_167002,Bishop_105,Peano_155129,Hausinger_030301,Bishop_100505}. In section \ref{Sec_CD} we focus on the lineshape of the cavity transmission in the nonlinear region. The experimental results are compared with predictions of a semiclassical theory. We find that good agreement can be obtained only in the limit of relatively small driving amplitudes. For higher driving amplitudes, better agreement is obtained with theoretical predictions derived by numerical integration of the master equation for the coupled system.
	
	\section{Experimental setup}
	
	The investigated device [see Fig.~\ref{FigNA}(a) and (b)] contains a CPW
	cavity resonator weakly coupled to two ports that are used for performing microwave
	transmission measurements [see Fig.~\ref{FigNA}(c)]. A persistent current flux qubit \cite{Mooij_1036},
	consisting of a superconducting loop interrupted by four Josephson junctions,
	is inductively coupled to the fundamental half-wavelength mode of the CPW resonator. We used a CPW line terminated by a low
	inductance shunt for qubit driving [see Fig.~\ref{FigNA}(b) and (c)]. We fabricated the device on a high resistivity silicon substrate in a two-step
	process. In the first step, the resonator and the control lines are defined
	using optical lithography, evaporation of a $190%
	\operatorname{nm}%
	$ thick aluminum layer and liftoff. In the second step, a bilayer resist is
	patterned by electron-beam lithography. Subsequently, shadow evaporation of
	two aluminum layers, $40%
	\operatorname{nm}%
	$ and $65%
	\operatorname{nm}%
	$ thick respectively, followed by liftoff define the qubit junctions. The chip
	is enclosed inside a copper package, which is cooled by a dilution
	refrigerator to a temperature of $T=$ 23  mK. We employed both passive and active shielding methods to suppress magnetic
	field noise. While passive shielding is performed using a three-layer high
	permeability metal, an active magnetic field compensation system placed
	outside the cryostat is used to actively reduce low-frequency magnetic field
	noise. We used a set of superconducting coils to apply DC magnetic flux. Qubit
	state control, which is employed in order to measure the qubit longitudinal
	$T_{1}$ and transverse $T_{2}$ relaxation times, is performed using shaped
	microwave pulses. Attenuators and filters are installed at different cooling
	stages along the transmission lines for qubit control and readout. A detailed
	description of sample fabrication and experimental setup can be found in
	\cite{Orgiazzi_104518,Bal_1324}.
	
	\section{The dispersive region}
	
	The circulating current states of the qubit are labeled as $\left\vert
	\curvearrowleft\right\rangle $ and $\left\vert \curvearrowright\right\rangle
	$. The coupling between the cavity mode and the qubit is described by the term
	$-g\left(  A+A^{\dag}\right)  \left(  \left\vert \curvearrowleft\right\rangle
	\left\langle \curvearrowleft\right\vert -\left\vert \curvearrowright
	\right\rangle \left\langle \curvearrowright\right\vert \right)  $ in the
	system Hamiltonian, where $A$ ($A^{\dag}$) is a cavity mode annihilation
	(creation) operator, and $g$ is the coupling coefficient. In the presence of
	an externally applied magnetic flux, the energy gap $\hbar\omega_{\mathrm{a}}$
	between the qubit ground state $\left\vert -\right\rangle $ and first excited
	state $\left\vert +\right\rangle $ is approximately given by $\hbar
	\omega_{\mathrm{a}}=\hbar\sqrt{\omega_{\mathrm{f}}^{2}+\omega_{\Delta}^{2}}$,
	where $\omega_{\mathrm{f}}=\left(  2I_{\mathrm{cc}}\Phi_{0}/\hbar\right)
	\left(  \Phi_{\mathrm{e}}/\Phi_{0}-1/2\right)  $, $I_{\mathrm{cc}}$
	($-I_{\mathrm{cc}}$) is the circulating current associated with the state
	$\left\vert \curvearrowright\right\rangle $ ($\left\vert \curvearrowleft
	\right\rangle $), $\Phi_{0}=h/2e$ is the flux quantum, $\Phi_{\mathrm{e}}$ is
	the externally applied magnetic flux and $\hbar\omega_{\Delta}$ is the qubit
	energy gap for the case where $\Phi_{\mathrm{e}}/\Phi_{0}=1/2$.
	
	In the dispersive region, i.e. when $g/\left\vert \Delta\right\vert \ll1$
	where $\Delta=\omega_{\mathrm{c}}-\omega_{\mathrm{a}}$ and $\omega_{\mathrm{c}}$ is the cavity mode angular frequency, the coupling between
	the cavity mode and the qubit gives rise to a resonance splitting. The steady
	state cavity mode response for the case where the qubit occupies the ground
	(first excited) states is found to be equivalent to the response of a mode
	having effective complex cavity angular resonance frequency $\Upsilon_{-}$
	($\Upsilon_{+}$), where $\Upsilon_{\pm}=\Upsilon_{\mathrm{c}}\pm
	\omega_{\mathrm{BS}}\pm\Upsilon_{\mathrm{ba}}$, $\Upsilon_{\mathrm{c}%
	}=\omega_{\mathrm{c}}-i\gamma_{\mathrm{c}}$ is the cavity mode intrinsic
	complex angular resonance frequency, with $\omega_{\mathrm{c}}$ being the angular
	resonance frequency, $\gamma_{\mathrm{c}}$ the linear damping rate, and
	$\omega_{\mathrm{BS}}=g_{1}^{2}/\left(  \omega_{\mathrm{c}}+\omega
	_{\mathrm{a}}\right)  $ the Bloch-Siegert shift \cite{Forn_237001}. The
	term $\Upsilon_{\mathrm{ba}}$ is given by
	\cite{Boissonneault_060305,Bishop_100505,Buks_033807,Xie_1806_05082}%
	\begin{equation}
	\Upsilon_{\mathrm{ba}}=-\frac{g_{1}^{2}}{\Delta}\frac{1-\frac{i}{\Delta T_{2}%
	}}{1+\frac{1}{\Delta^{2}T_{2}^{2}}+\frac{4g_{1}^{2}T_{1}E_{\mathrm{c}}}%
		{\Delta^{2}T_{2}}}\;, \label{Upsilon_ba bp}%
	\end{equation}
	where $g_{1}=g/\sqrt{1+\left(  \omega_{\mathrm{f}}/\omega_{\Delta}\right)
		^{2}}$ is the flux dependent effective coupling coefficient, $T_{1}=\gamma
	_{1}^{-1}$ and $T_{2}=\gamma_{2}^{-1}$ are the qubit longitudinal and
	transverse relaxation times, respectively, and $E_{\mathrm{c}}$ is the
	averaged number of photons occupying the cavity mode. Note that the imaginary
	part of $\Upsilon_{\pm}$ represents the effect of damping and the term
	proportional to $E_{\mathrm{c}}$ gives rise to nonlinearity. In the dispersive
	approximation this term is assumed to be small, i.e. $E_{\mathrm{c}}\ll
	\Delta^{2}T_{2}/4g_{1}^{2}T_{1}$. Note also that when $\Delta T_{2}\gg1$ the
	term $\Upsilon_{\mathrm{ba}}$ gives rise to a shift in the mode angular
	frequency approximately given by $\mp\chi$, where $\chi=g_{1}^{2}/\Delta
	$,\ and to a Kerr coefficient approximately given by $\pm\left(  g_{1}%
	^{4}/\Delta^{3}\right)  \left(  4T_{1}/T_{2}\right)  $.
	
	Network analyzer measurements of the cavity transmission are shown in
	Fig.~\ref{FigNA}(d). In the region where $\Delta>0$ (i.e. $\omega_{\mathrm{f}}<\sqrt{\omega_{\mathrm{c}}^{2}-\omega_{\Delta}^{2}}$) two peaks are seen in the
	cavity transmission, the upper one corresponds to the case where the qubit
	mainly occupies the ground state, whereas the lower one, which is weaker,
	corresponds to the case where the qubit mainly occupies the first excited state.
	
	\section{Qubit driving}
	
	\label{Sec_QD}
	
\begin{figure}
		[ptb]
		\begin{center}
			\includegraphics[
			height=2.7in,
			width=3.8in
			]%
			{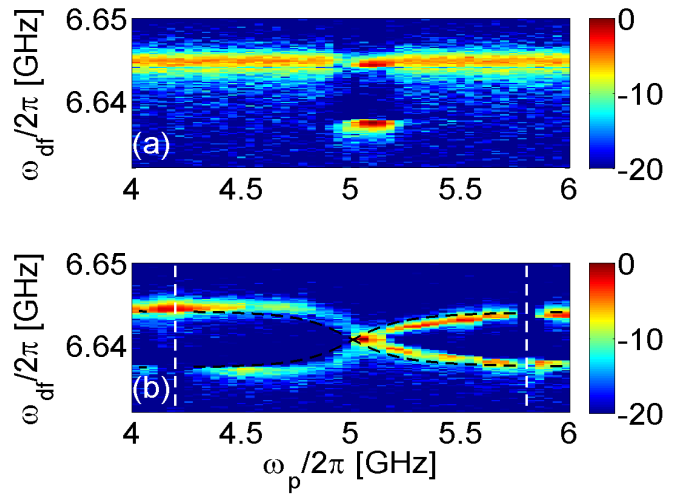}%
			\caption{The effect of qubit driving. Cavity transmission in dB units as a
				function of qubit driving frequency $\omega_{\mathrm{p}}/2\pi$ and cavity driving frequency $\omega_{\mathrm{df}}/2\pi$. The qubit driving amplitude $\omega_{1}$ in (b) is $100$ times larger compared with the values used in (a). For
				both plots the qubit frequency is given by $\omega_{\mathrm{a}}/2\pi=5\operatorname{GHz}$. The overlaid black dotted line in (b) is obtained by numerically calculating the transition frequencies between the lowest-lying eigenvalues of the Hamiltonian (\ref{H0}) using the following parameters $\omega_{1}/2\pi=0.5%
				\operatorname{GHz}%
				$, $\omega_{\mathrm{c}}/2\pi=6.6408%
				\operatorname{GHz}%
				$, $\omega_{\Delta}/2\pi=1.12%
				\operatorname{GHz}%
				$, $\omega_{\mathrm{f}}/2\pi=4.9%
				\operatorname{GHz}%
				$, $\omega_{\mathrm{a}}/2\pi=5.0%
				\operatorname{GHz}%
				$ and $g_{1}/2\pi=0.150%
				\operatorname{GHz}%
				$.}%
			\label{FigLNvsF}%
		\end{center}
	\end{figure}

	The flux qubit is driven by injecting a signal having angular frequency
	$\omega_{\mathrm{p}}$ and amplitude $\omega_{1}$ into the transmission line
	inductively coupled to the qubit [see Fig.~\ref{FigNA}(b) and (c)]. Network analyzer
	measurements of the cavity transmission as a function of $\omega_{\mathrm{p}}$
	for two fixed values of qubit driving amplitude $\omega_{1}$\ are shown in
	Fig.~\ref{FigLNvsF}(a) and (b). For both plots the qubit transition frequency
	is flux-tuned to the value $\omega_{\mathrm{a}}/2\pi=5%
	\operatorname{GHz}%
	$. The frequency separation between the two resonances that are shown in Fig.~\ref{FigLNvsF} is consistent with what is expected from the above-discussed dispersive shift $\mp\chi$, where $\chi=g_{1}^{2}/\Delta$. As can be seen from Fig.~\ref{FigLNvsF}, the visibility of the resonance
	corresponding to the qubit occupying the excited state at $6.637%
	\operatorname{GHz}%
	$ is affected by both angular frequency $\omega_{\mathrm{p}}$ and amplitude
	$\omega_{1}$ of qubit driving. These dependencies are attributed to
	driving-induced qubit depolarization.
	
	The comparison between Fig.~\ref{FigLNvsF}(a) and Fig.~\ref{FigLNvsF}(b),
	for which the qubit driving amplitude $\omega_{1}$ is $100$ times higher,
	reveals some nonlinear effects. One example is a nonlinear process of
	frequency mixing of the externally applied driving tones, which gives rise to pronounced features in the data when
	$\omega_{\mathrm{p}}$ is tuned close to the values $\left(  \omega
	_{\mathrm{a}}+\omega_{\mathrm{c}}\right)  /2=2\pi\times5.8%
	\operatorname{GHz}%
	$ and $\left(  3\omega_{\mathrm{a}}-\omega_{\mathrm{c}}\right)  /2=2\pi
	\times4.2%
	\operatorname{GHz}%
	$ [see the overlaid vertical white dotted lines in Fig.~\ref{FigLNvsF}(b)].
	
\begin{figure}
		[ptb]
		\begin{center}
			\includegraphics[
			height=3.8in,
			width=3.8in
			]%
			{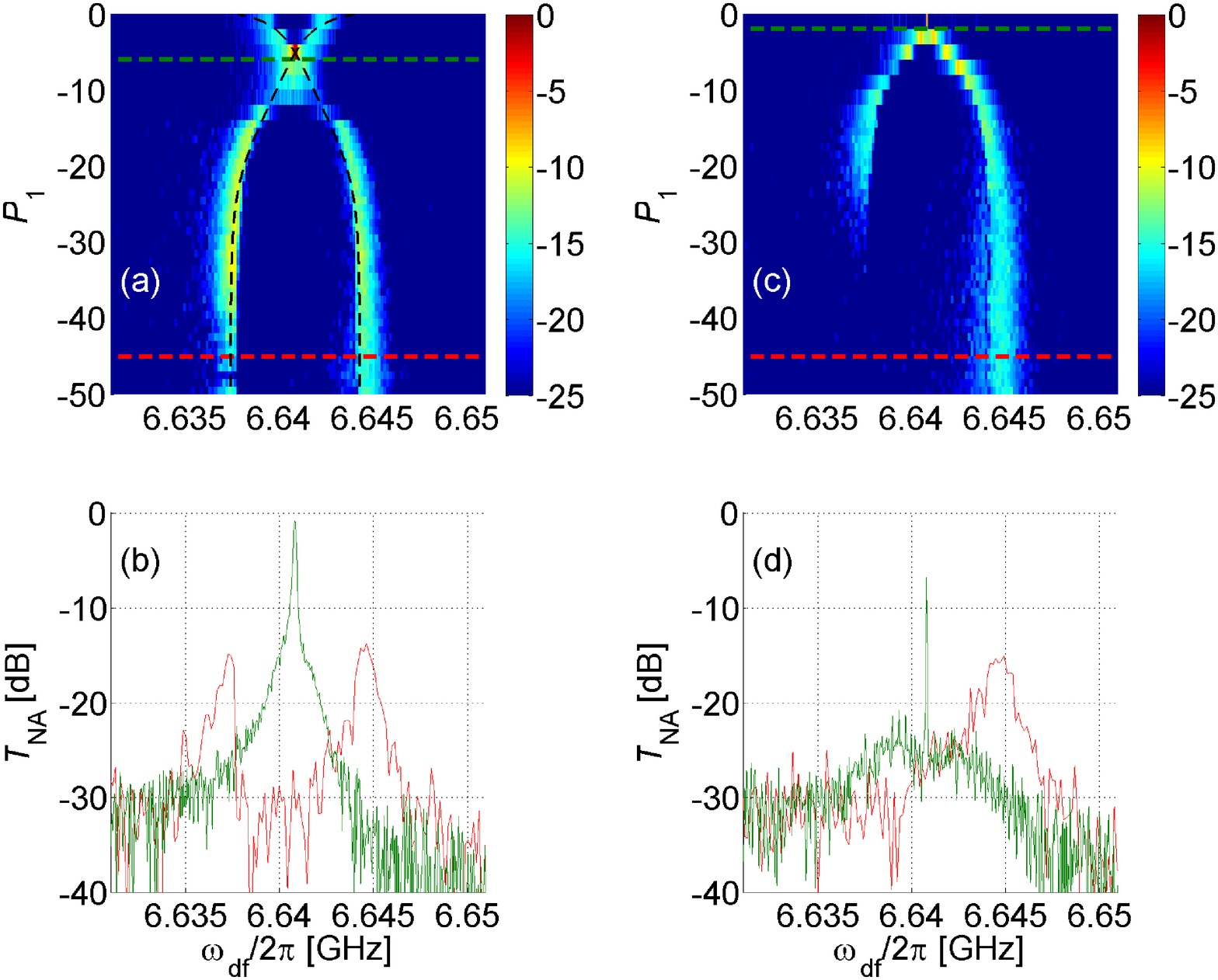}%
			\caption{
				Dependence on qubit driving amplitude. The driving frequency is $\omega_{\mathrm{p}}/2\pi=5.16%
				\operatorname{GHz}~%
				$ in (a) and (b) and $\omega_{\mathrm{p}}/2\pi=5.52%
				\operatorname{GHz}~%
				$ in (c) and (d). Cavity transmission in dB units as a function of cavity driving frequency $\omega_{\mathrm{df}}/2\pi$ and amplitude (in logarithmic scale) $P_{1}=20\log_{10}\left(  \omega_{1}%
				/\omega_{1,0}\right)  $ are shown in (a) and (c). Cross sections taken at values of $P_{1}$ indicated by colored horizontal dotted lines in (a) and (c) are shown using the corresponding colors in (b) and (d). The overlaid black dotted line in (a) is obtained by numerically calculating the transition frequencies between the eigenvalues of the Hamiltonian (\ref{H0}) using the following parameters $\omega_{\mathrm{p}}%
				/2\pi=5.16%
				\operatorname{GHz}%
				$, $\omega_{1,0}/2\pi=2.4%
				\operatorname{GHz}%
				$, $\omega_{\Delta}/2\pi=1.12%
				\operatorname{GHz}%
				$, $\omega_{\mathrm{f}}/2\pi=4.873%
				\operatorname{GHz}%
				$, $\omega_{\mathrm{a}}/2\pi=5.000%
				\operatorname{GHz}%
				$ and $g_{1}/2\pi=0.150%
				\operatorname{GHz}%
				$.
			}%
			\label{FigNA1}%
		\end{center}
	\end{figure}

	The dependence of cavity transmission on qubit driving amplitude
	$\omega_{1}$ with a fixed driving frequency of $\omega_{\mathrm{p}}/2\pi=5.16%
	\operatorname{GHz}~%
	[\omega_{\mathrm{p}}/2\pi=5.52%
	\operatorname{GHz}$%
	] is depicted in Fig.~\ref{FigNA1} (a) and (b) [(c) and (d)]. Cross sections of the color-coded
	plots shown in Fig.~\ref{FigNA1}(a) and (c) (corresponding to different values of the qubit driving amplitude $\omega_{1}$) are displayed in Fig.~\ref{FigNA1}(b) and (d).%
	
	As can be seen from the cross sections shown in Fig.~\ref{FigNA1}%
	(b) and (d), the cavity mode resonance line shapes exhibit hardening and softening
	effects (corresponding to positive and negative Kerr coefficient, respectively) in some region of the qubit driving amplitude $\omega_{1}$. Similar
	behavior is presented in Fig.~3 of Ref. \cite{Buks_033807}, which exhibits
	cavity mode resonance line shapes of the same device in the absence of qubit
	driving. However, while the nonlinearity observed in Ref. \cite{Buks_033807}
	is induced by cavity driving, the one shown in Fig.~\ref{FigNA1}(b) is
	induced by qubit driving.
	
	The cavity driving induced nonlinearity reported in Ref. \cite{Buks_033807} is
	well described by the above-discussed Kerr coefficients that can be calculated
	using Eq.~(\ref{Upsilon_ba bp}) [see also Eqs.~(4) and (A94) of Ref.
	\cite{Buks_033807}]. As is argued below, similar nonlinearity can be obtained
	due to qubit driving. In the rotating wave approximation (RWA) the Hamiltonian $\mathcal{H}_{0}$ of the closed system (consisting of a driven qubit and a coupled cavity mode) in a frame
	rotated at the qubit driving angular frequency $\omega_{\mathrm{p}}$ is found to be given
	by [see Eqs.~(\ref{eom a}), (\ref{eom p+'}) and (\ref{eom pz'}) of appendix \ref{App_RWA}]
	\cite{Mollow_2217,Agarwal_4555,Lewenstein_2048,Kowalewska_347,Zakrzewski_7717,Zhou_1515,Cohen1998atom}%
	\begin{equation}
	\hbar^{-1}\mathcal{H}_{0}=\left(  \omega_{\mathrm{R}}/2\right)  \Sigma
	_{z}^{\prime}-\Delta_{\mathrm{pc}}A^{\dag}A+\left(  g^{\prime}/2\right)
	\left(  A\Sigma_{-}^{\prime}+\Sigma_{+}^{\prime}A^{\dag}\right)  \;,
	\label{H_0 DS}%
	\end{equation}
	where $\omega_{\mathrm{R}}=\sqrt{\omega_{1}^{2}+4\Delta_{\mathrm{pa}}^{2}}$\ is
	the Rabi frequency, $\Delta_{\mathrm{pa}}=\omega_{\mathrm{p}}-\omega
	_{\mathrm{a}}$ and $\Delta_{\mathrm{pc}}=\omega_{\mathrm{p}}-\omega
	_{\mathrm{c}}$. The operators $\Sigma_{\pm}^{\prime}$ and $\Sigma_{z}^{\prime
	}$, which are defined by Eq.~(\ref{Ma T}), represent qubit operators in the
	basis of dressed states. This Hamiltonian (\ref{H_0 DS}) has the
	same structure as the Hamiltonian in the RWA of the same system in the absence
	of qubit driving [see Eq.~(A13) of Ref. \cite{Buks_033807}]. However, while
	for the case of no qubit driving the so-called crossing point occurs when the
	qubit angular frequency $\omega_{\mathrm{a}}$ coincides with the cavity mode
	angular frequency $\omega_{\mathrm{c}}$, the condition $\Delta_{\mathrm{pc}%
	}=\pm\omega_{\mathrm{R}}$ is satisfied at the crossing point of the driven
	system.
	
	Applying the transformation $\mathcal{H}_{\mathrm{T}}=U\mathcal{H}U^{\dag}$
	\cite{Boissonneault_060305} to the Hamiltonian $\mathcal{H}_{0}$ (\ref{H_0 DS}), where the
	unitary operator $U$ is given by $U=\exp\left(  \left(  \mathcal{N}%
	^{-1/2}\mathcal{S}/2\right)  \tan^{-1}\left(  g^{\prime}\mathcal{N}%
	^{1/2}/\Delta_{\mathrm{R}}\right)  \right)  $ and where $\Delta_{\mathrm{R}%
	}=\omega_{\mathrm{R}}+\Delta_{\mathrm{pc}}$, $\mathcal{N}=A^{\dag}A+\left(
	1+\Sigma_{z}^{\prime}\right)  /2$ and $\mathcal{S}=A\Sigma_{-}^{\prime
	}-A^{\dag}\Sigma_{+}^{\prime}$, yields (constant terms are disregarded)%
	\begin{align}
	\hbar^{-1}\mathcal{H}_{\mathrm{T}}  & =\left(  -\Delta_{\mathrm{pc}}+\xi
	\Sigma_{z}^{\prime}-\frac{g^{\prime4}\left(  1+A^{\dag}A\Sigma_{z}^{\prime
		}\right)  }{12\Delta_{\mathrm{R}}^{3}}\right)  A^{\dag}A\nonumber\\
	& +\frac{\left(  \omega_{\mathrm{R}}+\xi\right)  \Sigma_{z}^{\prime}}%
	{2}+O\left(  \left(  \frac{g^{\prime}}{\Delta_{\mathrm{R}}}\right)
	^{5}\right)  \;,\nonumber\\
	& \label{H_T}%
	\end{align}
	where $\xi=\left(  g^{\prime2}/\left(  4\Delta_{\mathrm{R}}\right)  \right)
	\left(  1-g^{\prime2}/\left(  3\Delta_{\mathrm{R}}^{2}\right)  \right)  $. Both hardening and softening effects are attributed to the term proportional to $A^{\dag}AA^{\dag}A\Sigma_{z}^{\prime}$ in Eq. (\ref{H_T}).
	
	The term proportional to $A^{\dag}A$ in the  Hamiltonian (\ref{H_T}) can be used to determine the shift in resonances that is induced by qubit driving. However, the comparison with the experimental results shown in Figs. \ref{FigLNvsF} and \ref{FigNA1} yields a moderate agreement. The inaccuracy is attributed to throwing away counter-rotating terms of the form $A \Sigma^\prime_+$ and $A^\dagger \Sigma^\prime_-$ in the derivation of the Hamiltonian (\ref{H_0 DS}). Much better agreement is obtained by numerically calculating the eigenvalues of the Hamiltonian (\ref{H0}). The results of this calculation are displayed in Figs. \ref{FigLNvsF}(b) and \ref{FigNA1}(a) by the overlaid black dotted lines. The parameters that have been assumed for the calculation are listed in the captions of Figs. \ref{FigLNvsF} and \ref{FigNA1}.
	
	As can be seen from both Fig.~\ref{FigLNvsF}(b) and the green-colored cross
	sections shown in Fig.~\ref{FigNA1}(b) and (d), in some regions of qubit driving parameters the two dressed fundamental resonances overlap. In the overlap region, a pronounced linewidth narrowing is observed [see Fig.~\ref{FigNA1}(b) and (d)].
	
	In appendix \ref{App_RWA} we show that the observed changes in linewidth of resonances can be qualitatively attributed to the Purcell effect for dressed states. In particular, both narrowing and broadening are demonstrated by Eqs. (\ref{Gamma_c}) and (\ref{Gamma_2}) below. However, the analytical expressions given by Eqs. (\ref{Gamma_c}) and (\ref{Gamma_2}), which have been obtained by assuming the limit of weak coupling and by employing the RWA, are not applicable in the region where the linewidth narrowing is experimentally observed. Consequently, direct comparison between theoretical predictions based on Eqs. (\ref{Gamma_c}) and (\ref{Gamma_2}) and data yields poor agreement. Moreover, the relatively intense driving in the region where the linewidth narrowing is experimentally observed gives rise to stochastic transitions between qubit states \cite{Gambetta_012112}. These stochastic transitions, which may give rise to the effect of motional narrowing \cite{Mukamel_1988,Li_1420}, cannot be adequately accounted for using the semicalssical approximation. Note that these phenomena are closely related to the effct of driving-induced spin decoupling (e.g. Fig. 7.27 of Ref. \cite{Slichter_Principles}).
	
	Numerical analysis based on the stochastic Schr{\"o}dinger equation is described below in appendix \ref{App_SLN}. We find that the effect of narrowing can be numerically reproduced provided that both qubit and cavity driving amplitudes are sufficiently large. The analysis in  this region is challenging, since relatively long integration times are needed to achieve conversion. As can be seen from Figs. \ref{FigNarrowingSim} and \ref{FigMetastableStates}, in the region were narrowing is numerically reproduced, the system becomes multistable.
		
	\section{Cavity driving}
	
	\label{Sec_CD}
	
	The nonlinear response of a microwave cavity coupled to a transmon superconducting qubit has recently been studied in Ref. \cite{Mavrogordatos_040402}. The experimental results, together with theoretical analysis \cite{Bishop_100505,Mavrogordatos_033828}, indicate that the response to strong cavity driving is affected by the significant coherent driving of the qubit as well as by the stochastic transitions between qubit states. The effect of cavity driving can be characterized by a dephasing rate and by a measurement rate. Both rates have been numerically calculated and analytically estimated in Ref. \cite{Gambetta_012112}.
	
\begin{figure}
		[ptb]
		\begin{center}
			\includegraphics[
			height=2.5958in,
			width=3.4546in
			]%
			{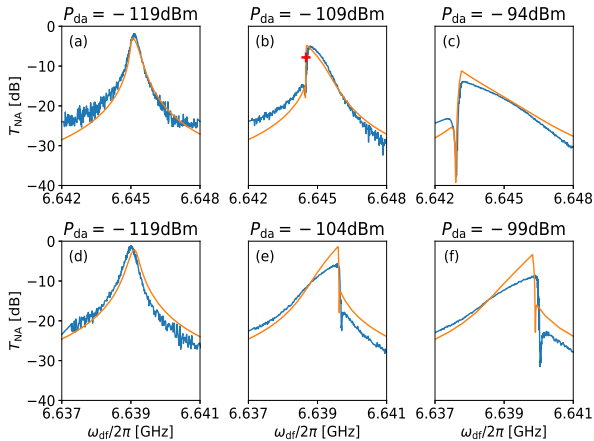}%
			\caption{Nonlinear response to cavity driving. The cavity transmission
				$T_{\mathrm{NA}}$ is measured as a function of cavity driving frequency
				$\omega_{\mathrm{df}}/2\pi$ for different values of the cavity driving power
				$P_{\mathrm{da}}$. These data are compared with a numerical calculation of the steady state of the Lindblad master equation (orange line), which is detailed in appendix \ref{App_SC}. In (a-c) the frequency $\omega_{\mathrm{f}}/2\pi$\ is flux-tuned to $5.5\operatorname{GHz}$, and in (d-f) to $7.8\operatorname{GHz} $. Different values of $\gamma_\mathrm{c}$ are used at each drive power to account for the increase in the quality factor of the cavity with occupation. We use $\gamma_\mathrm{c} = $ (a) $0.314$, (b) $0.251$, (c) $0.126$, (d) $0.314$, (e) $0.251$ and (f) $0.126\operatorname{MHz}$.
			}%
			\label{FigDip}%
		\end{center}
	\end{figure}

	Measurements of the cavity transmission $T_{\mathrm{NA}}$ of our device as a function of cavity driving frequency $\omega_{\mathrm{df}}/2\pi$ and power $P_{\mathrm{da}}$ are shown in Fig.~\ref{FigDip}. No qubit driving is applied during these measurements. We demonstrate nonlinearity of the softening type in Fig.~\ref{FigDip}(a-c), whereas hardening is demonstrated in Fig.~\ref{FigDip}(d-f). We obtained the data shown in Fig.~\ref{FigDip} by sweeping the cavity driving frequency $\omega_{\mathrm{df}}/2\pi$ upwards. Almost no hysteresis is observed when the sweeping direction is flipped.
	
	The measured cavity transmission $T_{\mathrm{NA}}$ can be compared with theoretical predictions based on the semiclassical approximation. Such a comparison has been performed in Ref. \cite{Buks_033807} based on data that has been obtained from the same device. Good quantitative agreement was found in the region of relatively small cavity driving amplitudes \cite{Buks_033807}.
	
\begin{figure}[ptb]
		\begin{center}
			\includegraphics[
			width=2.9656in
			]{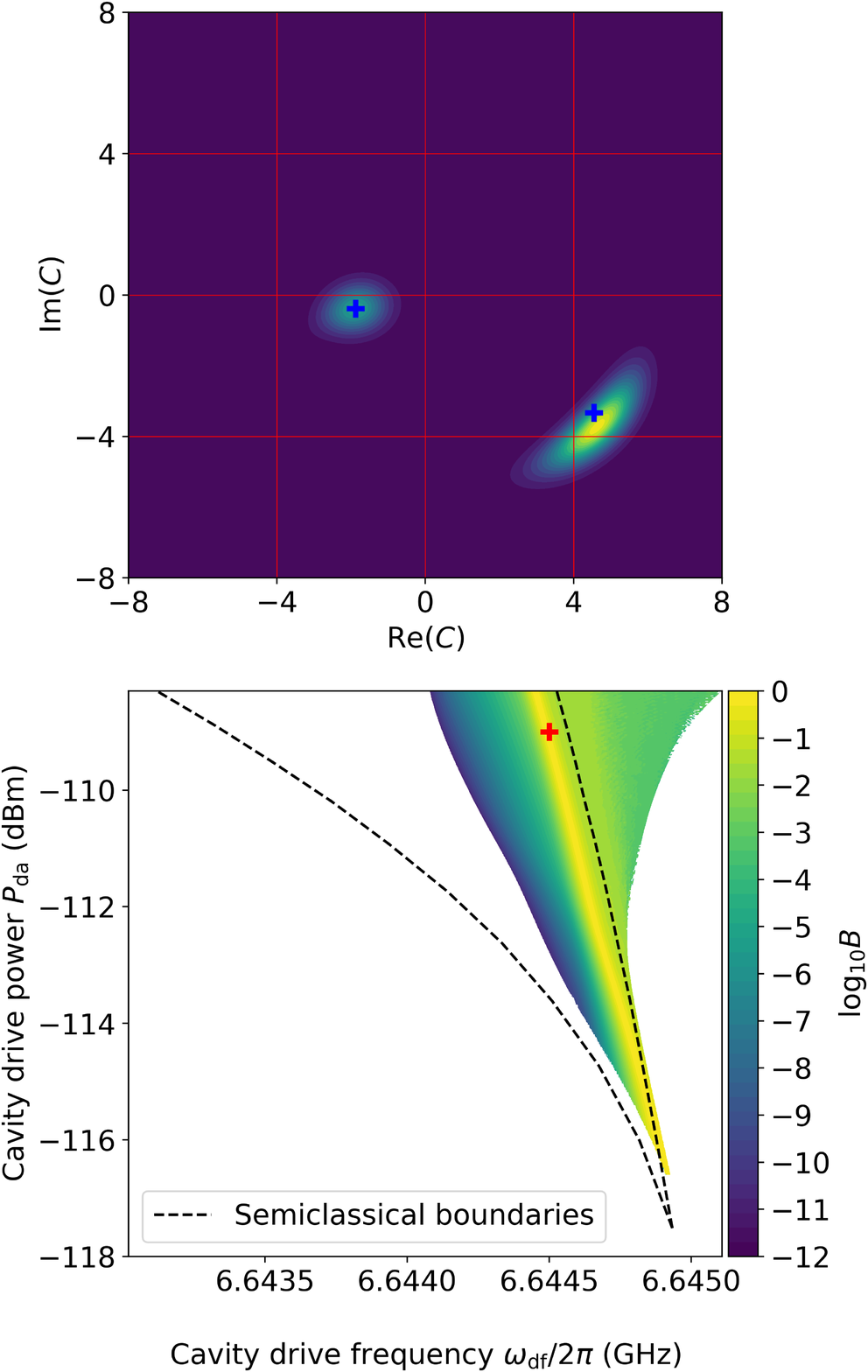}
		\end{center}
		\caption{The bistable regime. The cavity drive power is given by $P_{\mathrm{da}}=20\log_{10}\left(  \omega_{2}/\omega_{2,0}\right)  $ where $\omega_{2,0}/2\pi = 340 \operatorname{GHz}$. In (a) we plot the Wigner function of the cavity state in the bistable regime, which is obtained by solving for the steady state of the master equation at a drive power of $P_{\mathrm{da}} = -109 \text{dBm}$ and a drive frequency of $\omega_{\mathrm{da}} / 2 \pi= 6.6445 \text{GHz}$. These parameters are marked by the red crosses in panel (b) and in Fig. \ref{FigDip}(b). Two metastable states can be seen: a bright state at $C_\mathrm{b} = 4.64-3.73i$ and a dim state at $C_\mathrm{d} =-1.88-0.39i$. These two states correspond to the fixed points produced using the semiclassical equations of motion, marked by blue crosses. Next in (b) we examine the boundaries of the bistable regime. By examining the cavity Wigner function over a range of drive powers and frequencies we map the region in which we find two peaks corresponding to the bright and dim states. When two peaks can be identified we calculate the metric $B = 1 - \vert p_b - p_d \vert$ as a measure of bistability. This is plotted in the colormap above. Meanwhile the dashed black lines mark the boundaries of the region in which the semiclassical equations of motion have two fixed points. These methods produce significant overlap and both predict the onset of bistability around $P_\mathrm{da} = -117\mathrm{dB}$. We also see that the region of maximum bistability predicted by the master equation (yellow strip) lies either close to or within the semiclassical bistable region at all powers. However there are significant differences in the limits of the bistable region, particularly at the upper freuqency limit. The master equation predicts this limit should increase with drive power, whereas the semiclassical equations predict the opposite.}
		\label{FigWignerBistabilityMap}
	\end{figure}

	However, when the cavity is strongly driven, the nonlinearity introduced to the system by the qubit causes the onset of bistability and the semiclassical approximation alone is unable to reproduce the cavity transmission. This is because, despite accurately modeling the fixed points, which henceforth are referred to as the bright and dim metastable states (see Fig.~\ref{FigWignerBistabilityMap}), the semiclassical equations of motion give no information regarding the occupation probabilities of the two metastable states in the overall state of the system, which can be written as
	\begin{equation}
	\rho = p_\mathrm{b} \rho_\mathrm{b} + p_\mathrm{d} \rho_\mathrm{d} \;,
	\end{equation}
	where $\rho_\mathrm{b}(\rho_\mathrm{d})$ and $p_{\mathrm{b}}(p_{\mathrm{d}})$ represent the bright (dim) state and its probability respectively.
	
	The experimental results shown in Fig.~\ref{FigDip} exhibit a sharp dip in cavity transmission $T_{\mathrm{NA}}$ at drive powers above $-109\mathrm{dBm}$. A very similar feature has been experimentally observed before in \cite{Mavrogordatos_040402} and theoretically discussed in Refs. \cite{Bishop_100505,Mavrogordatos_033828}, for which the full quantum theory of the single nonlinear oscillator has been developed in \cite{Drummond_725}. The origin of this dip is the destructive interference between the two metastable states. Since the system is coupled to an external reservoir, fluctuations in the quantum state ensue and occasionally cause major switching events between the bright and dim states. When the complex amplitude of the cavity state is averaged over an ensemble of many such switching events, there is typically a narrow region in the frequency-power space where the two complex amplitudes partially cancel each other. By using the Lindblad master equation to model the system, we are able to take account of these fluctuations which cause these switching events and we produce the numerical fits seen in Fig. \ref{FigDip}. Comparison between the predictions derived from the numerical integration of the master equation and the ones analytically derived from the semiclassical equations of motion is shown in Fig. \ref{FigWignerBistabilityMap}.
	
	\section{Summary}
	
	Our main finding is the linewidth narrowing that is obtained by applying intense qubit driving. The effect is experimentally robust, however, its theoretical modeling is quite challenging. Further study is needed to explore the possibility of exploiting this effect for long-time storage of quantum information. We also find that bistability, which is predicted by the semiclassical model for monochromatic cavity driving, is experimentally inaccessible. This effect and related observations can be satisfactorily explained using numerical integration of the master equation for the coupled system. 
	
	\section{Acknowledgments}
	EB and PB contributed equally to this work. The work of the Waterloo group is supported by NSERC and CMC and the work of the Technion group by the Israeli Science Foundation.
	
	\appendix
	
	\section{Dressed states}
	
	\label{App_RWA} In this appendix the semiclassical dynamics of a driven qubit
	coupled to a cavity mode is discussed
	\cite{Mollow_2217,Agarwal_4555,Lewenstein_2048,Kowalewska_347,Zakrzewski_7717,Zhou_1515,Cohen1998atom}%
	. In the RWA the Hamiltonian $\mathcal{H}_{0}$ of the closed system in a frame
	rotated at the qubit driving angular frequency $\omega_{\mathrm{p}}$ is given
	by [see Eq.~(A13) of Ref. \cite{Buks_033807}]%
	\begin{align}
	\hbar^{-1}\mathcal{H}_{0}  &  =-\Delta_{\mathrm{pa}}\Sigma_{z}%
	+\frac{\omega_{1}\left(  \Sigma_{+}+\Sigma_{-}\right)  }{4}\nonumber\\
	&  -\Delta_{\mathrm{pc}}A^{\dag}A+\frac{g_{1}\left(  A\Sigma_{-}+\Sigma
		_{+}A^{\dag}\right)  }{2}\ ,\nonumber\\
	&  \label{H0}%
	\end{align}
	where $\Delta_{\mathrm{pa}}=\omega_{\mathrm{p}}-\omega_{\mathrm{a}}$,
	$\hbar\omega_{\mathrm{a}}$ is the qubit energy, $\Sigma_{z}$ is the qubit
	longitudinal operator, $\omega_{1}$ is the driving amplitude, $\Sigma_{+}$ and
	$\Sigma_{-}=\Sigma_{+}^{\dag}$ are qubit rotated transverse operators,
	$\Delta_{\mathrm{pc}}=\omega_{\mathrm{p}}-\omega_{\mathrm{c}}$, $\omega
	_{\mathrm{c}}$ is the cavity mode angular frequency, $A^{\dag}A$ is the cavity
	mode number operator and $g_{1}$ is the coupling coefficient.
	
	The Bloch equations of motion for the expectation values $a=\left\langle
	A\right\rangle $, $p_{+}=\left\langle \Sigma_{+}\right\rangle $ and
	$p_{z}=\left\langle \Sigma_{z}\right\rangle $ are obtained from the Heisenberg
	equations of motion and the commutation relations $\left[  A,A^{\dag}\right]
	=1$, $\left[  \Sigma_{z},\Sigma_{\pm}\right]  =\pm\Sigma_{\pm}$ and $\left[
	\Sigma_{+},\Sigma_{-}\right]  =2\Sigma_{z}$\ by adding fluctuation and
	dissipation terms and by averaging%
	\begin{align}
	\dot{a}  &  =-\left(  \gamma_{\mathrm{c}}-i\Delta_{\mathrm{pc}}\right)
	a-\frac{ig_{1}p_{+}}{2}\;,\label{a dot}\\
	\dot{p}_{+}  &  =-\left(  \gamma_{2}+i\Delta_{\mathrm{pa}}\right)
	p_{+}-\frac{i\omega_{1}p_{z}}{2}-ig_{1}ap_{z}\;,\label{p+ dot}\\
	\dot{p}_{z}  &  =-\gamma_{1}\left(  p_{z}-p_{0}\right)  +\frac{i\omega
		_{1}\left(  p_{+}^{\ast}-p_{+}\right)  }{4}+\frac{ig_{1}\left(  ap_{+}^{\ast
		}-p_{+}a^{\ast}\right)  }{2}\;, \label{pz dot}%
	\end{align}
	where overdot denotes time derivative, $\gamma_{\mathrm{c}}$ is the cavity
	mode damping rate, $\gamma_{1}$ and $\gamma_{2}$ are the qubit longitudinal
	and transverse damping rates, respectively, the coefficient $p_{0}%
	=-\tanh\left(  \left(  \hbar\omega_{\mathrm{a}}\right)  /\left(
	2k_{\mathrm{B}}T\right)  \right)  $ is the value of $p_{z}$ in thermal
	equilibrium (when $\omega_{1}=g_{1}=0$), $k_{\mathrm{B}}$ is the Boltzmann's
	constant and $T$ is the temperature. In the absence of coupling. i.e. when
	$g_{1}=0$, the steady state solution of Eqs.~(\ref{p+ dot}) and (\ref{pz dot})
	is given by $p_{+,\mathrm{ss}}=-\left(  i\omega_{1}p_{z,\mathrm{ss}}\right)
	/\left(  2\gamma_{2}+2i\Delta_{\mathrm{pa}}\right)  $ and $p_{z,\mathrm{ss}%
	}=p_{0}\left(  1+\left(  \gamma_{2}\omega_{1}^{2}\right)  /\left(  4\gamma
	_{1}\gamma_{2}^{2}+4\gamma_{1}\Delta_{\mathrm{pa}}^{2}\right)  \right)  ^{-1}$.
	
	Consider the transformation%
	\begin{equation}
	\left(
	\begin{array}
	[c]{c}%
	\Sigma_{+}^{\prime}\\
	\Sigma_{-}^{\prime}\\
	\Sigma_{z}^{\prime}%
	\end{array}
	\right)  =M_{\mathrm{a}}\left(
	\begin{array}
	[c]{c}%
	\Sigma_{+}\\
	\Sigma_{-}\\
	\Sigma_{z}%
	\end{array}
	\right)  \mathbf{\;,} \label{Ma T}%
	\end{equation}
	where%
	\begin{equation}
	M_{\mathrm{a}}=\left(
	\begin{array}
	[c]{ccc}%
	\frac{\sin\alpha+1}{2} & \frac{\sin\alpha-1}{2} & -\cos\alpha\\
	\frac{\sin\alpha-1}{2} & \frac{\sin\alpha+1}{2} & -\cos\alpha\\
	\frac{\cos\alpha}{2} & \frac{\cos\alpha}{2} & \sin\alpha
	\end{array}
	\right)  \mathbf{\;,} \label{Ma}%
	\end{equation}
	and where $\tan\alpha=\left(  -2\Delta_{\mathrm{pa}}/\omega_{1}\right)  $. Note
	that the transformed operators $\Sigma_{\pm}^{\prime}$ and $\Sigma_{z}%
	^{\prime}$ satisfy the commutation relations $\left[  \Sigma_{z}^{\prime
	},\Sigma_{\pm}^{\prime}\right]  =\pm\Sigma_{\pm}^{\prime}$ and $\left[
	\Sigma_{+}^{\prime},\Sigma_{-}^{\prime}\right]  =2\Sigma_{z}^{\prime}$
	provided that the original operators $\Sigma_{\pm}$ and $\Sigma_{z}$ satisfy
	$\left[  \Sigma_{z},\Sigma_{\pm}\right]  =\pm\Sigma_{\pm}$ and $\left[
	\Sigma_{+},\Sigma_{-}\right]  =2\Sigma_{z}$.
	
	Under this transformation the first two terms of the Hamiltonian
	$\mathcal{H}_{0}$ (\ref{H0}) become $-\Delta_{\mathrm{pa}}\Sigma_{z}%
	+\omega_{1}\left(  \Sigma_{+}+\Sigma_{-}\right)  /4=\left(  \omega
	_{\mathrm{R}}/2\right)  \Sigma_{z}^{\prime}$, where $\omega_{\mathrm{R}}%
	=\sqrt{\omega_{1}^{2}+4\Delta_{\mathrm{pa}}^{2}}$ is the Rabi frequency. In the
	RWA, in which counter-rotating terms are disregarded, the equations of motion
	(\ref{a dot}), (\ref{p+ dot}) and (\ref{pz dot}) are transformed into%
	\begin{align}
	\dot{a} &  =-\left(  \gamma_{\mathrm{c}}-i\Delta_{\mathrm{pc}}\right)
	a-\frac{ig_{1}^{\prime}p_{+}^{\prime}}{2}\;,\label{eom a}\\
	\dot{p}_{+}^{\prime} &  =-\left(  \gamma_{2}^{\prime}-\frac{i\omega
		_{\mathrm{R}}}{2}\right)  p_{+}^{\prime}-ig_{1}^{\prime}ap_{z}^{\prime
	}\;,\label{eom p+'}\\
	\dot{p}_{z}^{\prime} &  =-\gamma_{1}^{\prime}\left(  p_{z}^{\prime}%
	-p_{0}^{\prime}\right)  +\frac{ig_{1}^{\prime}\left(  ap_{+}^{\prime\ast
		}-p_{+}^{\prime}a^{\ast}\right)  }{2}\;,\label{eom pz'}%
	\end{align}
	where the effective coupling coefficient $g_{1}^{\prime}$ is given by
	$g_{1}^{\prime}=\left(  1/2\right)  \left(  \sin\alpha+1\right)  g_{1}$, the
	transformed damping rates $\gamma_{1}^{\prime}$ and $\gamma_{2}^{\prime}$ are
	given by $\gamma_{1}^{\prime}=\gamma_{2}+\left(  \gamma_{1}-\gamma_{2}\right)
	\sin^{2}\alpha$ and $\gamma_{2}^{\prime}=\left(  1/2\right)  \left(
	\gamma_{1}+\gamma_{2}\right)  +\left(  1/2\right)  \left(  \gamma_{2}%
	-\gamma_{1}\right)  \sin^{2}\alpha$, respectively, and the polarization
	coefficient $p_{0}^{\prime}$ is related to $p_{0}$ by
	\begin{equation}
	p_{0}^{\prime}=\frac{\sin\alpha}{\frac{\gamma_{2}}{\gamma_{1}}\left(
		1+\frac{\gamma_{1}-\gamma_{2}}{\gamma_{2}}\sin^{2}\alpha\right)  }%
	p_{0}\;.\label{p0'}%
	\end{equation}
	Note that the equations of motion (\ref{eom a}), (\ref{eom p+'}) and
	(\ref{eom pz'}) become unstable when
	\cite{Kocharovskaya_175,Hauss_037003,Hauss_095018,Andre_014016,Astafiev_840}%
	\begin{equation}
	g_{1}^{\prime2}\geq-\frac{2\gamma_{\mathrm{c}}\gamma_{2}^{\prime}}%
	{p_{0}^{\prime}}\left(  1+\frac{\Delta_{\mathrm{L}}^{2}}{\left(
		\gamma_{\mathrm{c}}+\gamma_{2}^{\prime}\right)  ^{2}}\right)  \;,
	\end{equation}
	where $\Delta_{\mathrm{L}}=\Delta_{\mathrm{pc}}-\omega_{\mathrm{R}}/2$.
	
	In the limit where the coupling coefficient $g_{1}^{\prime}$ is sufficiently
	small, at and near steady state the term $p_{z}^{\prime}$ in Eq.
	(\ref{eom p+'}) can be approximately treated as a constant, and consequently
	Eqs. (\ref{eom a}) and (\ref{eom p+'}) can be expressed in a matrix form as%
	\begin{equation}
	\frac{\mathrm{d}}{\mathrm{d}t}\left(
	\begin{array}
	[c]{c}%
	a\\
	p_{+}^{\prime}%
	\end{array}
	\right)  +M_{\mathrm{P}}\left(
	\begin{array}
	[c]{c}%
	a\\
	p_{+}^{\prime}%
	\end{array}
	\right)  =0\;,
	\end{equation}
	where the matrix $M_{\mathrm{P}}$ is given by%
	\begin{equation}
	M_{\mathrm{P}}=\left(
	\begin{array}
	[c]{cc}%
	\gamma_{\mathrm{c}}-i\Delta_{\mathrm{pc}} & \frac{ig^{\prime}}{2}\\
	ig^{\prime}p_{z}^{\prime} & \gamma_{2}^{\prime}-\frac{i\omega_{\mathrm{R}}}{2}%
	\end{array}
	\right)  \;.
	\end{equation}
	To lowest nonvanishing order in $g^{\prime}$ the eigenvalues of the matrix
	$M_{\mathrm{P}}$ are given by%
	\begin{align}
	\Gamma_{\mathrm{c}} &  =\gamma_{\mathrm{c}}-i\Delta_{\mathrm{pc}}+\frac
	{\frac{p_{z}^{\prime}g^{\prime2}}{2}}{\gamma_{2}^{\prime}-\gamma_{\mathrm{c}%
		}+i\Delta_{\mathrm{L}}}+O\left(  g^{\prime4}\right)  \;,\label{Gamma_c}\\
	\Gamma_{2} &  =\gamma_{2}^{\prime}-\frac{i\omega_{\mathrm{R}}}{2}-\frac
	{\frac{p_{z}^{\prime}g^{\prime2}}{2}}{\gamma_{2}^{\prime}-\gamma_{\mathrm{c}%
		}+i\Delta_{\mathrm{L}}}+O\left(  g^{\prime4}\right)  \;.\label{Gamma_2}%
	\end{align}
	The real parts of $\Gamma_{\mathrm{c}}$ ($\Gamma_{2}$) represents the
	effective damping rates $\gamma_{\mathrm{c,eff}}$ ($\gamma_{2\mathrm{,eff}%
	}^{\prime}$)\ of the cavity-like (qubit-like) mode. As can be seen from Eqs.
	(\ref{Gamma_c}) and (\ref{Gamma_2}), in this limit the coupling gives rise to
	repulsion-like behavior of the damping rates, i.e. $\left\vert \gamma
	_{\mathrm{c,eff}}-\gamma_{2\mathrm{,eff}}^{\prime}\right\vert >\left\vert
	\gamma_{\mathrm{c}}-\gamma_{2}^{\prime}\right\vert $ (it is assumed that
	$p_{z}^{\prime}<0$). This behavior can be considered as a generalization of
	the Purcell effect \cite{Purcell_839} for the case of dressed states.
	
	\section{Simulating line narrowing}
	\label{App_SLN}
	\begin{figure}
		[ptb]
		\begin{center}
			\includegraphics[
			width=3.4537in
			]
			{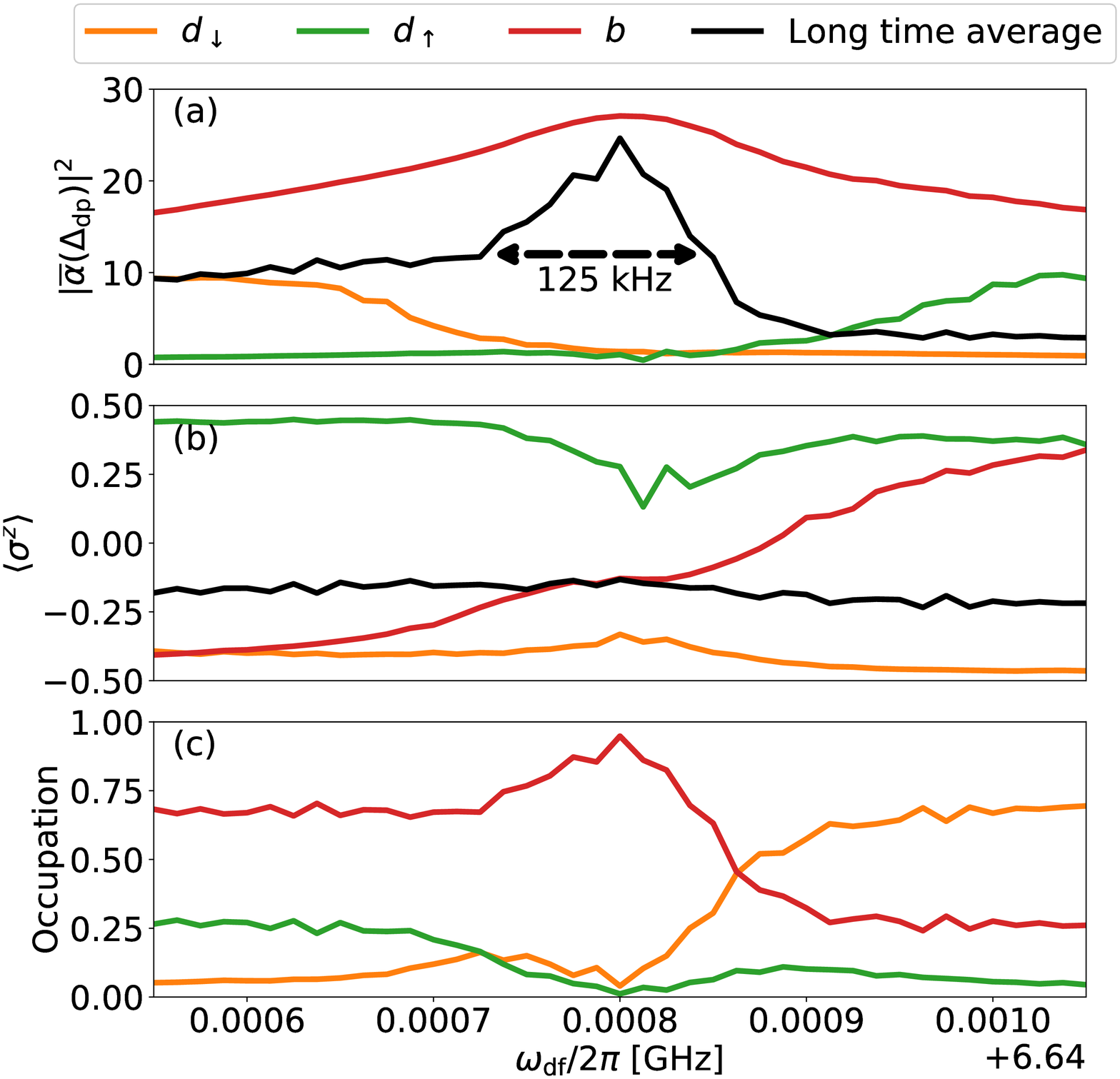}
			\caption{Simulation of the cavity spectrum in the nonlinear regime. We use a cavity drive amplitude of $\omega_2 / 2\pi = 1 ~\mathrm{MHz}$, a qubit drive frequency of $\omega_\mathrm{p} = 5.5 ~\mathrm{GHz}$ and a qubit drive amplitude of $\omega_1 = 0.863 ~\mathrm{GHz}$. The system displays multistability and three distinct metastable states can be identified, which are labelled by $d_\downarrow$, $d_\uparrow$ and $b$ and assigned the colours orange, green and red respectively. We plot the square cavity amplitude (a), qubit polarization (b) and occupation probability (d) of each of these three states against the cavity drive frequency $\omega_{\mathrm{df}} / 2 \pi$. In panel (a) we see that the cavity amplitude of state $b$ is significantly larger than the amplitudes of states $d_\downarrow$ and $d_\uparrow$. Hence we refer to $b$ as bright and $d_\downarrow$ and $d_\uparrow$ as dim. The black line is produced by averaging the cavity amplitude over $9.6~\mathrm{ms}$ of evolution before taking the square of the absolute value. It displays a narrow resonance at the bare cavity frequency. The full width at half maximum is only $125 ~\mathrm{kHz}$, which is $33\%$ of the natural linewidth of $377 ~\mathrm{kHz}$. In panel (b) we see that states $d_\downarrow$ and $d_\uparrow$ occur when the qubit is polarized up and down respectively whereas the qubit polarization associated with state $b$ varies with the drive frequency. Finally in panel (c) we see the occupation probabilities of the three states. Away from the cavity resonance the stability of state $b$ falls. This causes the narrowing observed in panel (a).}
			\label{FigNarrowingSim}
		\end{center}
	\end{figure}
	
	\begin{figure*}
		[ptb]
		\begin{center}
			\includegraphics[height=2.5949in]%
			{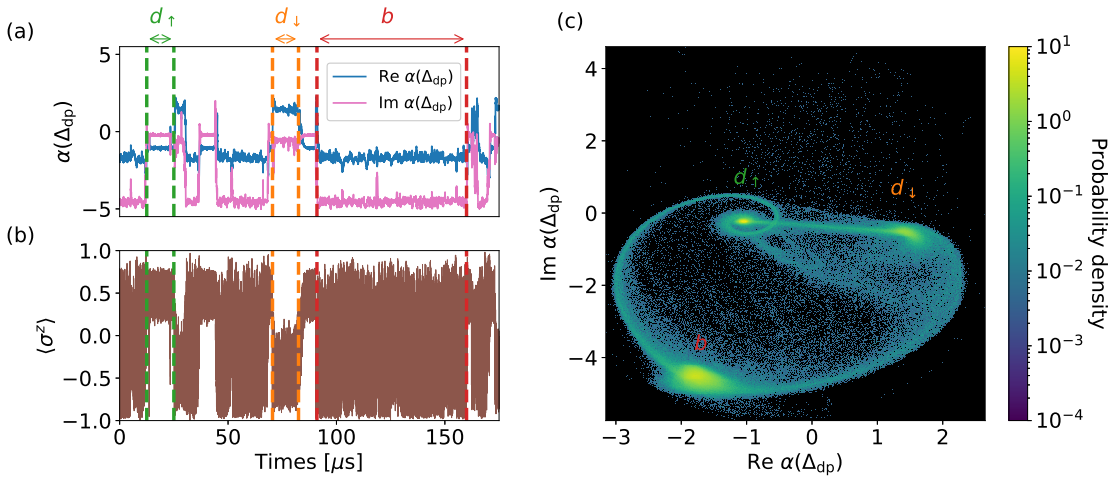}%
			\caption{Here we examine a quantum state trajectory produced at $\omega_{\mathrm{df}} / 2 \pi = 6.6409 ~ \mathrm{GHz}$. In panel (a) we plot the real and imaginary parts of the cavity amplitude. The cavity is observed to jump between three metastable states. Examples of states $d_\uparrow$, $d_\downarrow$ and $b$ are highlighted between the vertical dashed lines in green, orange and red respectively. By referring to panel (b) we see that state $d_\uparrow$ occurs when the qubit has positive polarization, state $d_\downarrow$ occurs when it has negative polarization and state $b$ occurs when the qubit freely varies over the range $-1 < \langle \sigma^z \rangle < 1$. In panel (c) we plot a histogram of the cavity amplitude throughout $9.6 ~\mathrm{ms}$ of evolution. The three metastable states are clearly identified as three clusters in the plane. Switching pathways leading between these clusters can also be observed.}
			\label{FigMetastableStates}
		\end{center}
	\end{figure*}

	Experimentally we have observed narrowing in the cavity spectrum which occurs when a drive is applied to the qubit. In an attempt to model this narrowing we perform simulations of the cavity response by unravelling the Lindblad master equation using a quantum jump (Monte-Carlo) stochastic Schro{\"o}dinger equation. In a frame rotating with the qubit drive at angular frequency $\omega_{\mathrm{p}}$ we use the rotating wave approximation (RWA) to write down the Hamiltonian as:
	\begin{equation}
	\hbar^{-1} \mathcal{H}(t) = \hbar^{-1} \mathcal{H}_{0} + \omega_{2} \big( A \exp(i \Delta_{\mathrm{dp}} t) + A^\dagger \exp(-i \Delta_{\mathrm{dp}} t) \big)
	\end{equation}
	where the time independent part of the Hamiltonian $\mathcal{H}_{0}$ is given in Eq.~(\ref{H0}) and the time dependent cavity drive oscillates at the frequency $\Delta_{\mathrm{dp}} = \omega_{\mathrm{df}} - \omega_{\mathrm{p}}$. In order to describe dissipation due to loss of photons from the cavity we use the Lindblad operator $\sqrt{\gamma_{\mathrm{c}}} \, A$, while to describe dissipation in the qubit we use $\sqrt{\gamma} \, \Sigma^-$. After combining these elements the evolution of the state of the system is described by:
	\begin{equation}
	\partial_t \rho = -\frac{i}{\hbar} [\mathcal{H}(t),\rho] + \gamma_c \mathcal{D}[A]\rho + \gamma_1 \mathcal{D}[\Sigma^+]\rho.
	\end{equation}
	
	We use this equation to numerically evolve the state $\rho(t)$ over time and study the cavity amplitude $\langle A \rangle = \Tr ( \rho(t) A)$. We find that the time-dependence of $\langle A \rangle$ contains two main frequencies: $\Delta_{\mathrm{dp}}$ and $\Delta_{\mathrm{cp}}$, due to the drive and cavity frequency respectively. The experimental data presented in Fig.~\ref{FigNA1} were measured by mixing the signal transmitted through the cavity with a reference at the cavity drive frequency. Therefore in order to model the transmitted power $T_{\mathrm{NA}}$ we must examine the cavity amplitude $\langle A \rangle$ in a frame rotating with the drive. This is given by
\begin{equation}
\alpha(\Delta_{\mathrm{dp}},t) =  \, \Tr \Big(\rho(t) A \Big) \exp \big( - i \Delta_{\mathrm{dp}} t  \big).
\end{equation}
Input-output relations can then be used to calculate $T_{\mathrm{NA}}$ from this amplitude.
	
	We now attempt to reproduce the spectrum seen in Fig.~\ref{FigNA1}(b). In order to observe narrowing we must drive the cavity in the nonlinear regime. We take a cavity drive amplitude of $\omega_2 / 2 \pi = 1.00 ~\mathrm{MHz}$. The remaining parameters are set to $\omega_1 / 2 \pi = 1.726 ~\mathrm{MHz}$, $\omega_{\mathrm{p}} / 2 \pi = 5.50 ~ \mathrm{GHz}$, $g_{1} / 2 \pi = 0.150 ~ \mathrm{GHz}$, $\omega_{\mathrm{a}} / 2 \pi = 5 ~ \mathrm{GHz}$, $\gamma_{c} / 2 \pi = 377 ~ \mathrm{kHz}$ and $\gamma_{1} / 2 \pi = 40.7 ~ \mathrm{kHz}$. Using these parameters we produce the spectrum in Fig. \ref{FigNarrowingSim} by evolving the state of our system over $9.6 ~\mathrm{ms}$ for a range of cavity drive frequencies. The long time average $\overline{\alpha}(\Delta_{\mathrm{dp}})$ displays a full width at half maximum of $125 ~\mathrm{kHz}$, significantly less than the natural linewidth of $377 ~\mathrm{kHz}$.

	This narrowing can be explained when we realise that in the presence of a strong cavity drive the system displays multistability and the line narrowing is due to a bright cavity state ($b$) which is most stable over a narrow range of frequencies close to the bare cavity resonance. Close to the cavity resonance the system occupies the bright state and the transmitted power is high. However away from this point the system may also occupy two other dim states ($d_\downarrow$ and $d_\uparrow$), which causes a sharp drop in the transmitted power and a narrow linewidth.
	
	In Fig. \ref{FigMetastableStates} we examine these metastable states more closely. We plot the cavity amplitude and qubit polarization over $170 ~\mathrm{\mu s}$ of evolution at $\omega_d/ 2 \pi = 6.6409 ~\mathrm{GHz}$. The two dim states, labelled $d_{\uparrow}$ and $d_{\downarrow}$, occur when the qubit is polarized in the up and down directions respectively. Meanwhile the bright state occurs when the qubit is depolarised and varies widely over the range $-1 < \langle \sigma^z \rangle < 1$.
	
	\section{Spectra calculations}
	\label{App_SC}
	In order to calculate the response of our system to cavity driving (without qubit driving) we use the
	following master equation:
	\begin{align}
	\partial_{t} \rho=  &  -i [\mathcal{H}/\hbar, \rho] + \gamma_{c}
	\mathcal{D}[A]\rho+ \gamma_{1} \mathcal{D}[\Sigma_{+}] \rho \;,
	\label{master_equation}%
	\end{align}
	which consists of non-unitary components calculated according to
	$\mathcal{D}(L) \rho= L^{\dagger}\rho L - \frac{1}{2}(L^{\dagger}L \rho+ \rho
	L^{\dagger}L)$ and a unitary component which obeys the Hamiltonian given by
	\begin{align}
	\hbar^{-1}\mathcal{H}  &  =-\Delta_{\mathrm{da}}\Sigma_{z}%
	+ \omega_{2}\left(  A^{\dagger}+ A \right) \nonumber\\
	&  -\Delta_{\mathrm{dc}}A^{\dag}A+ \frac{g_{1}}{2}\left(  \Sigma_{-} A + \Sigma
	_{+}A^{\dag}\right).
	\end{align}
	In the above the detuning between the cavity drive and the cavity resonance is
	given by $\Delta_{\mathrm{dc}} = \omega_{\mathrm{df}} -
	\omega_{\mathrm{c}}$, while the detuning between the cavity drive and the qubit
	frequency is given by $\Delta_{\mathrm{da}} = \omega_{\mathrm{df}} - \omega
	_{\mathrm{a}}$. Since the relaxation rate of the qubit depends on the
	magnetic field detuning from the symmetry point we must take account of this
	in our calculations. For Figs. \ref{FigDip}(a-c) we have $\omega_{\mathrm{f}}
	/ 2 \pi= 5.5 \mathrm{GHz}$ and $\gamma_{1} / 2\pi = 6.29 \mathrm{kHz}$, whereas for Figs. \ref{FigDip}(d-f) we have $\omega_{\mathrm{f}} / 2 \pi= 7.8
	\mathrm{GHz}$ and $\gamma_{1} / 2\pi = 4.02 \mathrm{kHz}$.
	
	The master equation above does not include a Lindblad operator to describe pure dephasing of the flux qubit. Since we are operating the qubit far from its symmetry point, pure dephasing will be dominated by flux noise, and in \cite{Orgiazzi_104518} the power spectral density (PSD) of this noise was found to have a $1/f^{0.9}$ form. Unfortunatley we cannot account for this noise in the master equation, because the Markovian approximation requires that the PSD is well-behaved at zero frequency. However, even without the inclusion of pure dephasing, the master equation is still able to explain the major features of the spectra measured in Fig. \ref{FigDip}.

\newpage
\bibliographystyle{ieeepes}
\bibliography{acompat,Eyal_Bib}

\end{document}